\documentstyle{article}
\begin{document}
\begin{center}
\null
\rightline {UFR-HEP/04-99}
\rightline {LPHEA/99-02}
\vskip1truecm

\bf\Large On the confining potential in 4d $SU(N_c)$ \\ gauge theory with dilaton
\vskip2truecm
\bf\large M. Chabab$^{1,}$\footnote[2]{mchabab@ucam.ac.ma}, R. Markazi$^1$ and
E.H. Saidi\footnote[1]{H-saidi@fsr.ac.ma}\\

\small $^1$UFR-Physique des Hautes Energies, Facult\'e des \small
Sciences, Av. Ibn Battouta\\ \small B.
P. 1014 , Rabat, Morocco.\\
\small $^2$LPHEA, D\'epartement de Physique, Facult\'e des
Sciences-Semlalia, B.P. 2390\\ \small Marrakech, Morocco.\\
\end{center}
\vskip1truecm
\begin{abstract}
Using the formal analogy between the Dick superstring inspired model of ref.[6] and
the problem  of building of Eguchi Hanson metric in 4d N=2 harmonic superspace (hs),
we derive a general formula for the quark-quark interaction potential V(r) including
the Dick confining potential. The interquark potential V(r) depends on the
dilaton-gluon coupling and may be related to the parameterization of confinement by
the quark and gluon vacuum condensates. It is also shown how the axion field may be
incorporated in agreement with 10d type IIB superstring requirements. Others features
are also discussed.
\end{abstract}

\newpage
\section{Introduction}

\qquad The dilaton $\phi$ and the axion $\chi$ are scalar fields predicted  by
superstring theory [1]. Both of them arise in a natural way in the massless spectrum
of 10 dimensional (10d) type IIB superstring theory [1,2]and its lower dimensional
compactifications. In the language of 4d gauge theory of field strength  $F_{\mu\nu}$
and its dual $\tilde{F}_{\mu\nu}$, the $\phi$ and $\chi$ have very special couplings.
The dilaton $\phi$ couples to gauge fields through a term $exp(\phi)F^2$ and the axion
$\chi$ couples to  the topological term. The $\phi$ and $\chi$ fields play a central
role in superstring dualities [3], F-theory compactifications [4] and in the
derivation of the exact results in 4d N=2 supersymmetric gauge theories [5].\\
Recently it was observed in [6] that a string inspired coupling of a dilaton
 $\phi$ to 4d  $SU(N_c)$ gauge fields $A_{\mu}=T^{a}A_{\mu}^{a}$, with $T^a$ the
 $(N_c^2-1)$ $SU(N_c)$ generators, yields a phenomenologically interesting potential
 V(r) for the quark-quark
interactions. Following [6,7], this potential is obtained as follows: First we start from
 the following model for the scalar field-gluon coupling
\begin{equation}
 L(\phi,A)=-{1\over 4G(\phi)} F_{\mu \nu}^{a}F_{a}^{\mu \nu}-{1\over2}(\partial_{\mu}{\phi})^2+W(\phi)+J_{\mu}^{a}A_{a}^{\mu}\qquad.
\label{eq:quantum}
\end{equation}
Then we choose $G(\phi)$, the coupling of the scalar field $\phi$ to the $SU(N_c)$ field strength $F_{\mu\nu}$, and the interacting lagrangian $W(\phi)$ as:\\
$$G(\phi)=const.+{f^2\over \phi^2}\qquad $$
\begin{equation}
W(\phi)={1\over 2}m^2\phi^2 \qquad \label{eq:quantum}
\end{equation}
\\
where the parameter $f$  is a scale characterizing the strength of the scalar-gluon coupling and
 m is the mass of the scalar field $\phi$. Next we consider the equations of motion of the
  fields $A_\mu$ and $\phi$  and solve them for static point like color source of current
  density $J_a^{\mu}=\rho_a\eta^{\mu 0}$. After some straightforward algebra, we find that the
  Dick quark interaction potential $V_{D}(r)$ is given by:
\begin{equation}
V_D(r)={1\over r}-f\sqrt{N_c\over{2(N_c-1)}}\ln[exp(2mr)-1]\qquad
\label{eq:quantum}
\end{equation}
Eq.(3) is very remarkable since for large values of $r$ it leads
to a confining potential
$V_D(r)\sim2fm\sqrt{N_c\over{2(N_c-1)}}r$. In this regard, we will
show in this paper that for a general gluon-dilaton coupling
$G(\phi)$, the quark interactions potential $V(r)$ reads as:
\begin{equation}
V(r)= \int dr{G[\phi (r)]\over {r^2}}\qquad .
\label{eq:quantum}
\end{equation}
Such form of the potential is very attractive. On one hand it extends the usual
Coulomb formula $V_c \sim 1/r$ which is recovered from eq.(4) by taking  $G=1$.
Moreover for $G \sim r^2$, which by the way corresponds to a coupling
$G(\phi)\sim~\phi^{-2}$, and $W(\phi)={m^2\over 2}\phi^2$,  $m \ne 0$, eq.(4) yields a
linearly increasing interquark potential $V\sim r$ having the good behaviour to
describe the $SU(N_c)$ quarks confinement [6,7,8]. On the other hand eq.(4) may be
also used to describe other non perturbative effects associated with higher dimension
quark and gluon vacuum condensates. Following [8], see also [9], one may extract
interesting phenomenological informations on the dilaton-gluon coupling $G[\phi]$ by
comparing eq.(4) to the Bian-Huang-Shen's potential $V_{BHS}(r)$ namely:
\begin{equation}
V_{BHS}(r)\sim {1\over r}-{\sum_{n\ge 0}C_nr^n}\qquad \label{eq:quantum}
\end{equation}
where $C_n$'s are related to the quark and gluon vacuum condensates. In fact one can
do better if one can put the coupling $G(\phi)$ in the form $G[\phi(r)]$. In this case
one can predict the type of vacuum condensates  of the $SU(N_c)$ gauge theory which
contributes to the quark-quark interaction potential [9]. Thus, although, the
derivation of the formula (4) for the interquark potential from eq.(1) is by itself an
important result, there remain however other steps to overcome before one can exploit
eq.(4). As mentioned above, a crucial step is to determine what type of couplings
$G(\phi)$ which can be put in the form $G[\phi(r)]$. In other words for what couplings
$G(\phi)$, one can solve the equation of motion of the scalar field $\phi$. This is a
technical problem without solving it one cannot get V(r). An other step which remains
to clarify is to show how the effective model (1) may be got from a more fundamental
theory. If this is possible, one may for instance justify the mass scale $f$
introduced by hand in eqs.(2,3). One might also get some informations on the axion
field couplings and more generally on the moduli of 10d  superstrings compactified on
six dimensional compact manifolds and especially type IIB on Calabi-Yau Threefolds
[10]. In trying to explore eq.(4), we have observed some remarkable facts among which
we quote the three following :\\ \qquad 1) The functional $G[\phi(r)]$, and then the
potential V(r) of eq.(4) may be obtained from the following one dimensional lagrangian
$L_D$
\begin{equation}
L_D={1\over 2}(y')^2+r^2W(y/r)+{\alpha\over {2r^2}}G(y/r)\qquad \label{eq:quantum}
\end{equation}
where $y=r\phi$, $y'=({dy\over dr})$and $\alpha={g^2\over {16\pi^2}}{{N_c-1}\over{2N}}$ and where $g$ is the gluon coupling constant. In particular  $L_D$  reads, for $W(\phi)$ and $G(\phi)$  like in eq(2),as:
\begin{equation}
2L_D=(y')^2+m^2y^2+{\mu^2\over {y^2}}\qquad \label{eq:quantum}
\end{equation}
where $\mu= \alpha f^2$.\\ \qquad 2) Eq.(7) has a striking resemblance with the
following harmonic superspace lagrangian $L_{EH}$ used in [11] in the derivation of
the 4d Eguchi-Hanson metric

\begin{equation}
2L_{EH}=(D^{++}\omega)^2+{m^{++}}^2\omega^2+{{\mu^{++}}^2\over {\omega^2}}\qquad .
\label{eq:quantum}
\end{equation}
In this equation, $\omega$ is an analytic harmonic superspace (hs) superfield taken to
be dimensionless, $ D^{++}$ is the hs covariant derivative and $m^{++}$  and
$\mu^{++}$  are coupling  constants. More details on hs tools will be described in
section 3. Much more precision can be found in [12]. For the moment note only the
formal analogy between y, dy/dr, m and $\mu$   of eq.(7) with $\omega$,
$(D^{++}\omega)$, $\mu^{++}$  and $m^{++}$ respectively. Both of models eqs.(7) and
(8), involve hermitian fields with a self interacting potential proportional to the
inverse of the square of the scalar field variable.\\ \qquad 3) The Dick potential (3)
is viable only for non zero mass dilaton field exactly as in 4d N=2  supersymmetric
theories where the scalar potential is proportional to the mass eigenvalues of the
central charges of the 4d N=2 superalgebra [13,5]. Recall by the way that in 4d N=2
supersymmetric QFT, mass terms are generated by central charges. We shall see in
sections 3 and 4 that this formal analogy between the Dick model (1) and 4d N=2 QFT's
is much deeper since it allows us to derive a new model containing eq.(1) and where
the symmetries behind the solvability of Dick equations as well as the couplings of
both the dilaton and axion fields are manifest.

 The aim of this paper is to generalize the Dick model (1)  by exploiting the formal analogy
with 4d N=2 supersymmetric theories formulated in hs [11] and using known 4d N=2 exact
results. In addition to the derivation of new model exhibiting a U(1) gauge
invariance, we give an interpretation of the mass scale $f$, introduced by hand in
eq.(2), as a Kahler moduli of a blown up SU(2) singularity of Calabi-Yau threefold of
type II superstring compactifications. The appearance of the local U(1) symmetry  in
the analysis of eqs.(1-3) has a quite interesting consequence as it offers a
possibility to incorporate in the game the axion field $\chi$ couplings. Recall that
in Dick model as formulated in [6], the role of the topological field $\chi$ is
ignored. We shall show in section 4 how this field can be incorporated by going in a
general gauge other than $\phi=\phi^*$.

The presentation of this paper is as follows: In section 2, we formulate the Dick problem as a
one dimensional field theory. In section 3 we give general solutions including those of [6,7].
In section 4, we review briefly the building of the Eguchi Hanson hyperkahler metric in
harmonic superspace. In section 5 we use the formal analogy between the Eguchi Hanson model and
our one dimensional field theoretical formulation of the Dick problem to determine the dilaton couplings, the axion ones and interprete the mass scale $f$ as a kind of  Fayet-Iliopoulos coupling. Our conclusion is given in section 6.\\
\section{The Dick model as a one dimensional field theory.}

\qquad Following ref.[6,7], the analysis of the Coulomb problem of the theory (1) is
based on considering a point like static color source which in its rest frame is
described  by a current $J_a^\mu=g\delta(r)C_a\eta_0^\mu$ where $C_a$ is the
expectation value of the $SU(N_c)$ generator for a normalized spinor in the color
space. These $C_a$'s satisfy the algebraic identity
\begin{equation}
\sum_{a=1}^{N^2-1}C_a^2={(N_c-1)\over 2N_c}
\label{eq:quantum}
\end{equation}
The next step is to use the residual SO(3) space symmetry, which remains after setting $J_a^\mu=\rho_a\eta_0^\mu$, to rewrite the equations of motion
$$[D_{\mu},G^{-1}(\phi)F^{\mu\nu}]=J^\nu $$
\begin{equation}
\partial_\mu \partial^\mu \phi={\partial W\over \partial \phi}- {1\over 4}F_{\mu \nu}^a F_a^{\mu\nu}{\partial G^{-1}(\phi)\over \partial \phi}
\label{eq:quantum}
\end{equation}
into a simple form. Indeed setting $F_a^{0i}=-{gC_a\over 4\pi}\partial_i V$; $\alpha={g^2\over 16\pi}{(N_c-1)\over 2N_c}$ one finds after some easy algebra:
$${dV\over dr}=r^{-2}G[\phi]\qquad \qquad \qquad (a)$$
\begin{equation}
\label{eq:quantum}
\end{equation}
$$\Delta\phi={\partial W\over \partial\phi}+{\alpha\over r^4}{\partial G(\phi)\over
\partial\phi}\qquad \qquad (b)$$ Note that eqs.(11) have four unknown field
quantities; the  field $\phi$, the interacting color potential V(r), the dilaton-gluon
coupling $G(\phi)$  and  the $\phi$ potential $W$. To solve eqs.(11) one has to fix
two of them. For example choosing $2W=m\phi^2$and $G(\phi)$ as in equation (3), one
finds :
\begin{equation}
\phi=\phi_D(r)=r^{-1}{[{\alpha f\over m}(1-exp(-2mr)]}^{1/2}
\label{eq:quantum}
\end{equation}
\begin{equation}
V_D(r)={1\over r}-f\sqrt{N_c\over {2(N_c-1)}}ln[exp(2mr)-1]\qquad .
\label{eq:quantum}
\end{equation}
In general given $G(\phi)$   and $W(\phi)$, the color potential
V(r) can be exactly determined up on  solving one equation namely
equation (11-b). For later use, let us introduce the new
dimensionless field  $y= r\phi$  and take the spherical coordinate
frame $(r,\theta,\varphi)$ to rewrite the lagrangian (1) as:
\begin{equation}
L=-{r^2\over {2G(\phi)}}F_{0r}^aF_a^{0r}-{r^2\over 2}\partial_r\phi\partial^r\phi+r^2W(\phi)+F_{0r}^a\rho_a\qquad .
\label{eq:quantum}
\end{equation}
In deriving eq.(14), we have used  the stationarity  of the color source, the SO(3)
symmetry and the identity $\Delta(1/r)=\delta(r)$. Putting this equality back into
eq.(14) and using the change of  variable  $y=r\phi$  together with the convention
notations $y'=\partial_ry$; $\partial^r\phi=r^2\partial_r \phi$ as well as eq.(9), one
gets the lagrangian form eq.(6). Consequently the coupling $G(\phi)$ of eq.(1) appears
as a part of interacting  potential of the one dimensional field theory eq.(6). From
this point of view, the finding of the interquark potential V(r) is equivalent to
solve the equation of motion\\
\begin{equation}
y''{\partial L_D\over {\partial y'}}+y'{\partial L_D\over {\partial y}}+\partial_r^{exp}L_D=0\qquad.
\label{eq:quantum}
\end{equation}
\section{Solving the Dick model}
\qquad First of all observe that the lagrangian (6) including the Dick model (7) is a
particular one dimensional field theory of lagrangian
\begin{equation}
L={1\over 2}(y')^2-U(y,r)
\label{eq:quantum}
\end{equation}
where $U(y,r)$ is a priori an arbitrary potential. Though simple, this theory is not
easy to solve except in some special cases. A class of solvable models is given by
potentials of the form :
\begin{equation}
U(y)= \lambda^2 y^{2(n+p)}+\gamma^2 y^{2(q-n)}+\delta y^k
\label{eq:quantum}
\end{equation}
where $n, p, q$ and $k$ are numbers and $\lambda^2$ , $\gamma^2$ and $\delta$ are
coupling constants scaling as $(lenght)^{-2}$. The next thing to note is that eq.(17)
has no explicit dependence in $r$ and consequently the following  identity usually
hold :
\begin{equation}
y'^2 = U+c
\label{eq:quantum}
\end{equation}
where c is  a constant. Actually eq.(18) is just an integral of motion which may be
solved  under some assumptions. Indeed by making appropriate choices of  the coupling
$\lambda$ as well as the integral constant  c , one may linearise y' in eq.(18) as
follows :
\begin{equation}
y' = U_1+U_2\qquad.
\label{eq:quantum}
\end{equation}
Once the linearisation in y' is achieved and the terms $U_1$ and $U_2$ are identified,
we can show that the solutions of eq.(18) are classified by the product $U_1U_2$ and
the ratio $U_1/U_2$. In what follows we discuss briefly some interesting examples. For
convenience let us rewrite eq.(18) as:
\begin{equation}
y'^2 = w_0 + w_1 + C_0
\label{eq:quantum}
\end{equation}
where $w_0=m^2y^2 $  and  $w_1$  is the interaction term which we take for the moment
to be the Dick interaction that is $w_1=c_1^2y^{-2}$, where $ c_1 $ is a coupling
constant. Starting from eq.(20), it is not difficult to see that there are two
possibilities to put it in the form (19):
\subsection{First possibility: the Dick solution}
This corresponds to take $w_0=U_1^2$    that is $U_1=my$ and $U_2=c_1y^{-1}$. Putting
back into eq.(19) one gets the Dick solution given by eqs.(12,13).
\subsection{Second possibility: New solutions}
In this case the mass term  is related to the product $U_1U_2$  as
:
\begin{equation}
U_1U_2=\pm {1\over 2}m^2y^2
\label{eq:quantum}
\end{equation}
Eq.(21) cannot however determine $ U_1 $ and $U_2$ independently as in general the
following realizations are all of them candidates,
\begin{equation}
U_1=\lambda y^{n+p};\qquad\qquad  U_2=\gamma y^{q-n} \label{eq:quantum}
\end{equation}
where the integers $p$ and $q$ are such that $p+q =2$ and where $\lambda \gamma=\pm m^2$. A remarkable example corresponds to take $p+q=1$. In this case we distinguish two solutions according to the sign of the product of $\lambda \gamma$ . For $\lambda \gamma=+m^2$, the solution is
\begin{equation}
y(r)=[{1\over \lambda} tan({nmr\over \sqrt2}+const)]^{1\over n}\qquad .
\label{eq:quantum}
\end{equation}
For $\lambda\gamma=-m^2$ , we have:
\begin{equation}
y(r)=[-{1\over \lambda} tanh({nmr\over \sqrt2}+const)]^{1\over n}\qquad.
\label{eq:quantum}
\end{equation}
The solutions (23) and (24) have quite interesting features inherited essentially from
the features of  tan  and  tanh  functions. We remark that for n=0 the solution is:
\begin{equation}
y(r)=const.exp({{\lambda+\gamma}\over \sqrt2}r)\qquad. \label{eq:quantum}
\end{equation}
In the end of this section, it should be noted that one can go beyond the above
mentioned solutions which are just special case of general models involving
interactions classified by the following constraint equations
\begin{equation}
U_1.U_2\sim y^k
\label{eq:quantum}
\end{equation}
where $U_1$ and $U_2$ are as in eq.(19) and $k$ is an integer. For $k=0$, one gets the
Dick model and for $k=2$ one has solutions described in subsection 3-2 . For general
values of  $k $, one has to know moreover the ratio $U_1/U_2$ in order to work out
solutions. For the example where
$$ U_1=\lambda y$$
\begin{equation}
U_2=\gamma y^{k-1}\qquad;\quad k\quad integer
\label{eq:quantum}
\end{equation}
one can check, after some straightforward algebra, that the solution of y is just a
generalization of eq.(12) that is
\begin{equation}
y_k(r)=[r\phi_D]^{2\over (2-k)}\qquad .
\label{eq:quantum}
\end{equation}
For $k = 0$, one discovers the solution (12).
\section{The Eguchi Hanson hs model}

\qquad To start recall that Eguchi Hanson metric is a vacuum
solution of the self dual euclidean four dimensional gravity. It
is a Ricci flat hyperkahler  metric having an $SU(2)\times U(1)$
isometry. There are different, but equivalent, ways of writing
this metric. A remarkable way of expressing this  metric is that
using a local coordinate system   exhibiting manifestly the $
SU(2)\times U(1)$ symmetry. The element of length $ds^2$ reads as:
\begin{equation}
ds^2=g_{iajb}df_1^{ia}df_1^{jb}+k_{iajb}df_2^{ia}df_2^{jb}
+h_{iajb}df_2^{ia}df_1^{jb}
\label{eq:quantum}
\end{equation}
where the metric factors are given by:
$$g_{iajb}=\epsilon_{ab}\epsilon_{ij}-{4f_{2ia}f_{2jb}\over
{f_1^{kc}f_{1kc}+f_2^{kc}f_{2kc}}}\qquad\qquad (a)$$
\begin{equation}
k_{iajb}= \epsilon_{ab}\epsilon_{ij}-{4f_{1ia}f_{1jb}\over
{f_1^{kc}f_{1kc}+f_2^{kc}f_{2kc}}}\qquad\qquad (b)
\label{quantum}
\end{equation}
$$h_{iajb}=-{4f_{1ia}f_{2jb}\over {f_1^{kc}f_{1kc}+f_2^{kc}f_{2kc}}}\qquad\qquad\qquad
c)$$ together with the SU(2) isovector constraint
\begin{equation}
\epsilon_{ab}(f_1^{ia}f_2^{jb}+f_1^{ja}f_2^{ib})-\lambda^{ij}=0\qquad.
\label{quantum}
\end{equation}
A tricky way to derive this metric is to use results of 4d N=2 supersymmetric non linear $\sigma$ models. In the harmonic superspace approach where 4d N=2 supersymmetry is manifest, the field theoretical model giving the family of Eguchi Hanson metrics reads in the superfield language as :
\begin{equation}
S[\omega]={1\over {2k^2}}\int dz^{(-4)}du[(D^{++}\omega)^2-{m^{++2}}\omega^2-{{\lambda^{++2}}\over {\omega^2}}]
\label{quantum}
\end{equation}
In this equation $\omega=\omega (x_A,\theta ^+,\bar{\theta}^+,u)$,
is an analytic hs superfield taken to be dimensionless.
$D^{++}=(u^{+i}{\partial\over {\partial u^{-i}}}-2\theta ^+ \sigma
^m \bar{\theta}^+\partial _m)$ is the hs covariant derivative;
$dz^{-4}$ is the analytic superspace measure with U(1) Cartan
charge (-4) and the couplings $ m^{++}$ and $ \lambda^{++}$ are
given by
\begin{equation}
m^{++}=u_i^+u_j^+m^{ij}\qquad;\lambda^{++}=u_i^+u_j^+\lambda^{ij}
\label{quantum}
\end{equation}
where $ u_i^+$ and $u_i^-$ are the harmonic variables parameterizing the
$SU(2)/U(1)\approx S^2$ sphere. We shall not use here after these hs tools, we are
only interested in the formal analogy with the Dick problem. This is why we shall give
here after only the necessary material. For more details on the HS method and the
derivation of the Eguchi Hanson metric see [11]. Note also that the Eguchi Hanson
metric with $SU(2)\times U(1)$ isometry corresponds to $ m^{++}=0$. Metrics with $
m^{++} \ne 0 $ have a $U(1)\times U(1)$ symmetry and fall in the family of multicenter
metrics [14,15].
    Let us take $ m^{++}=0$  and sketch the main steps in putting eq.(32) in the form (29-31).
In fact there are  two possible paths  one  may  follow: First, a direct method which
consists to start from the superfield equation of motion of the hermitian hs
superfield $\omega$:
\begin{equation}
D^{++2}\omega={\lambda^{++2}\over {\omega^3}}
\label{quantum}
\end{equation}
and use the $\theta-$expansion of the superfield $\omega$, that is
\begin{equation}
\omega=\phi+\theta^{+2}M^{(-2)}+\bar{\theta}^{+2}\bar{N}^{(-2)}+\theta ^+ \partial^m\bar{\theta}^+B_m^{(-2)} +
\theta^{+2}\bar{\theta}^{+2}P^{(-4)};
\label{quantum}
\end{equation}
where we have ignored fermions. Then fix N =2  supersymmetry partially on shell by
eliminating the auxiliary fields $P^{(-4)}$ and $B_m^{(-2)}$. The relevant equations
are those corresponding to the projection of eq.(34) along the $\theta^+=0$ and
$\theta ^+\sigma^m\bar{\theta}^+$ directions, i.e.
$$\partial^{++2}\phi=\lambda^{++2}/\phi^3$$
\begin{equation}
\partial^{++}B_m^{-2}=2(\partial_m-{3\over 2}{\lambda^{++2}\over {\phi^4}}B_m^{(-2)})\phi\qquad.
\label{quantum}
\end{equation}
The next thing  to do is to find the explicit dependence of $\phi$ and $B_m^{(-2)}$ in
harmonic variables $u_i^\pm $  by solving eqs.(36). Then put the solution into eq.(32)
once the integrations with respect to $\theta^+$ and $\bar{\theta}^+$  are performed.
In other words put the solutions $\phi=\phi(u_i^\pm) $
,$B_m^{(-2)}=B_m^{(-2)}(u_i^\pm)$ into the following component field action:
\begin{equation}
S[\omega]\sim {1\over k^2}\int dx^4 du[\partial^{++}B_m^{(-2)}\partial^m\phi+\partial^mB_m^{(-2)}\partial^{++}\phi]\qquad .
\label{quantum}
\end{equation}
The last step is to integrate with respect to harmonic variables. Once this is done, we get the bosonic part of the 4d N=2 supersymmetric non linear $\sigma $ model from which one can read the Eguchi Hanson metric in the $\omega$ representation.
    The second method, which interest us here, is indirect but it has the merit of being based on hs  superfield theory exhibiting manifestly the $SU(2)\times U(1)$ symmetry. The main steps of this approach are as follows:

1. Instead of working with a real superfield $\omega$, we take a
complex superfield $\omega$ : $\bar{\omega}\ne \omega$ .

2. Modify the action eq.(32) as:
\begin{equation}
S[\omega]\sim {1\over {2k^2}}\int dz^{(-4)} du[|(D^{++}+iV^{++})\omega|^2+\lambda^{++}V^{++}]\qquad,
\label{quantum}
\end{equation}
where $V^{++}$ is a U(1) gauge superfield. Eq.(38) is invariant under the following
gauge transformations of parameter $\lambda$.
\begin{equation}
\omega'=\exp(-i\lambda)\omega\quad;\qquad V'^{++}= V^{++} + D^{++} \lambda\qquad.
\label{quantum}
\end{equation}

Note that $V^{++}$ has no kinetic term. It is an auxiliary superfield which can be eliminated
through its equation of motion namely
\begin{equation}
2V^{++}={1\over {\omega \bar{\omega}}}[i(\bar{\omega}D^{++}\omega-\omega D^{++}\bar{\omega})-\lambda^{++}].
\label{quantum}
\end{equation}
For the special case where $\omega$ is real; $\bar{\omega}=\omega$ , eq.(40) reduces
to
\begin{equation}
V^{++}=-\lambda^{++}/\omega^2
\label{quantum}
\end{equation}
and consequently the action (38) coincides with eq.(32). Note by the way that the term
$\lambda^{++}V^{++}$  is a Fayet-Iliopoulos (FI) coupling.

3. Rewrite eq.(38) in an equivalent form by using O(2) notations i.e. express the
complex supefield  $\omega=\omega_1+i\omega_2$ as an O(2) doublet $(\omega_1,
\omega_2)$ and introducing two other auxiliary superfield $ F_1^{++}$  and $ F_2^{++}$

\begin{equation}
S[\omega_1,\omega_2,F_1^{++},F_2^{++}]={1\over {2k^2}}\int dz^{(-4)}du \lbrack(F_1^{++})^2+2F_1^{++}D^{++}\omega_1+(1\leftrightarrow 2)
\label{quantum}
\end{equation}
$$\qquad\qquad\qquad -V^{++}(\omega_1F_2^{++}-\omega_2F_1^{++}+\lambda^{++})\rbrack$$

Eliminating $V^{++}$,$F_1^{++}$   and $F_2^{++}$  and choosing the gauge $\omega_2=0$ we reproduce the action $ S_{EH}$ (32) with  $m^{++}=0$. The second order action(42) is interesting since it has a manifest SU(2) invariance rotating $\omega_i$ and $F_i^{++}$  . To make this invariance more explicit we make the following change for both $(\omega_1,F_1^{++})$     and $(\omega_2,F_2^{++})$.
$$\omega=U_a^-q^{+a};\qquad F^{++}=U_a^+q^{+a}$$
\begin{equation}
q^{+a}=\epsilon^{ab}q_b^+;\qquad q_a^+=(q^+,\bar{q}^+);\qquad \epsilon^{12}=1.
\label{quantum}
\end{equation}
Thus for both $\omega_1$ and$\omega_2$, we have $\omega=U_a^-q_I^{+a}$ with $I=1,2 $
and so  on. Putting back into eq.(42), we get the following  action,
\begin{equation}
S={1\over {2k^2}}\int dz^{(-4)}
du[\bar{q}_1^+D^{++}q_{1}^++\bar{q}_2^+D^{++}q_{2}^+++V^{++}(\bar{q}_1^+q_2^++\bar{q}_2^+q_1^++\lambda^{++})]\qquad.
\label{quantum}
\end{equation}
This action has the invariance under the following groups:

        i) O(2) gauge group acting  as
\begin{equation}
\delta q_I^+=\epsilon_{IJ}\lambda q_J^+;\qquad\delta V^{++}=D^{++}\lambda ;\qquad \bar{\lambda}=\lambda\qquad.
\label{quantum}
\end{equation}

        ii) U(1) subgroup of the rigid SU(2) automorphism group of supersymmetry that leaves $\lambda^{++}$ invariant.

        iii) The SU(2) Pauli Cursey symmetry rotating $q_I^+$  and $\epsilon_{IJ}\bar{q}_J^+$.\\
    Now starting from the last form of the E.H. action (42) and solving the $\theta^+=0$ and $\theta^+ \sigma^m \bar{\theta}^+$components of the equations of motion:
$$D^{++}q_I-\epsilon_{IJ}V^{++}q_J^+=0 $$
\begin{equation}
\epsilon^{IJ}\bar{q}_I^+q_J^++\lambda^{++}=0
\label{quantum}
\end{equation}
in the Wess-Zumino gauge, one gets by following the same lines described for the
direct method, the E.H. metric (29-31).\\

\section{The Dick model revisited}

\qquad In section 2 we have learnt that the Dick problem may be formulated as a one
dimensional field theory of lagrangian  $L_D$ given by eq.(16) namely:
\begin{equation}
2L_D=({dy\over dr})^2-m^2y^2-{\mu^2\over y^2}
\label{quantum}
\end{equation}
    In section 3 we have showed that hyperkhaler metrics of the Eguchi Hanson family can be
derived from the following  4d  N=2 supersymmetric model,
\begin{equation}
2L_{EH}=(D^{++}\omega)^2-{m^{++}}^2\omega^2-{{\mu^{++}}^2\over \omega^2}
\label{quantum}
\end{equation}
    In section 4 we have seen that this lagrangian is equivalent to the following first order one, once the auxiliary U(1) gauge superfield $V^{++}$ is eliminated through its equation of motion:
\begin{equation}
2L'_{EH}=|{D^{++}\omega}|^2-{m^{++}}^2\omega
\bar{\omega}-V^{++}(\bar{\omega}D^{++}\omega - \omega D^{++}
\bar{\omega}- \mu^{++})+{V^{++}}^2\omega \bar{\omega}
\label{quantum}
\end{equation}

This form of $L_{EH}$ may be also transformed into two other forms
as shown on eq.(42) and eq.(44). The difference between $L_{EH}$
and $L'_{EH} $ is that in eq.(48) $\omega $ is hermitian whereas
in eq.(49) $\omega$ is complex. As we have seen, we can go from
eq.(49) to eq.(48) either by constraining the  superfield to be
real, that is :
\begin{equation}
\omega_2=0,
\label{quantum}
\end{equation}
 Or equivalently by keeping $\omega_2\ne 0 $ and working in the Wess-Zumino
 gauge:
\begin{equation}
D^{++}V^{++}=0 \label{quantum}
\end{equation}
Eq.(51) turns on  to be helpful in the derivation of the Eguchi Hanson metric. Now
using the formal analogy between eqs.(47) and (48), it is not difficult to see that
the $L_D $ lagrangian may be also formulated in terms of the auxiliary fields $ F $
and $\bar{F}$ as:
\begin{equation}
L_2=F\bar{F}+\bar{F}\partial y+F\partial \bar{y}+V[\xi+i(y\bar{F}-\bar{y}F)]
\label{quantum}
\end{equation}
where V is a one dimensional U(1) gauge field and $\xi$ is a 1d constant vector  breaking explicitly invariance under space translations. Note that in this formulation, the two scalars $y_1$ and $y_2$ of the complex field $y={1\over \sqrt{2}}(y_1+iy_2)$  represent respectively the dilaton $\phi$ and the axion $\chi$ in agreement  with the requirement of F-theory and 10d type IIB superstring and 4d N supersymmetric gauge theory. Eliminating the auxiliary fields F and   through their equations of motion namely
$$F=-(\partial +iV)y=-\nabla y\qquad\qquad(a) $$
\begin{equation}
\label{quantum}
\end{equation}
$$\bar{F}=-(\partial +iV)\bar{y}=-\nabla \bar{y}\qquad \qquad (a)$$ one obtains the
following first order lagrangian $ L_1$
\begin{equation}
L_1=|\nabla y|^2+m^2y\bar{y}+\xi V\qquad.
\label{quantum}
\end{equation}
Moreover eliminating the auxiliary U(1) gauge field V through its equation of motion
\begin{equation}
V ={-1\over {2y\bar{y}}}[\xi +i(\bar{y}\partial y+ y \partial \bar{y})],
\label{quantum}
\end{equation}
one gets the one dimensional field theory of the dilaton-axion system extending eq.(7)
which may be recovered from eqs.(54,55) by going to the gauge fixing $y_2=\chi=0$.
However to exhibit the effect of the axion field $\chi$, one has to keep $ \chi \ne 0
$ and imposes a constraint on the gauge field $V$ that we write as follows:
\begin{equation}
C(V,\partial V)=0 \label{quantum}
\end{equation}
    Using this constraint, the first order lagrangian is no longer invariant  under the change
$ y\rightarrow e^{-i\psi}y $ and  $ V \rightarrow V+\partial \psi
$, where $\psi$ is the U(1) gauge parameter, but as a counterpart
one can work out a non trivial solution for the axion field $\chi
=\chi (r) $  by solving the conjugate where the field  V should be
substituted in equations of motion (55) $\nabla^2y=m^2y $ and its
complex by the value $ V_0(r) $ verifying the constraint eq.(56)
and satisfying the identity $\partial (\bar{y}yV_0)=0 $. In the
end of this study we would like to note that the term $\xi V $
appearing in eq.(52) plays a similar role as the Fayet Iliopoulos
term $ m^{++}V^{++}$ of the 4d N=2 supersymmetric Eguchi Hanson
model (32) . Thus the mass scale  $f$  introduced by hand in the
Dick model may be viewed, under some assumptions, as the scale of
breaking of the U(1) symmetry rotating the dilaton and axion
fields. Recall by the way that in general supersymmetric gauge
theories with a U(1) gauge invariance, the FI term is generally
used to break supersymmetry and/or gauge invariance. The FI
couplings are Kahler moduli of Calabi-Yau threefolds on which 10d
type II superstrings are compactified  on, and their magnitude are
of order of the Calabi-Yau compactification scale.\\
\\
\section{Conclusion}

\qquad Inspired from the dilaton-gluon coupling in superstring
theory, Dick built a field theoretical model having the remarkable
property  of leading to a confining quark-quark interaction
potential. The model is mainly a 4d $SU(N_c)$ gauge theory coupled
to a massive scalar field $\phi$ of lagrangian (1) and a
dilaton-gluon coupling $G(\phi)=1+f^2/\phi^2 $, where $f$  is a
mass scale introduced by hand. The parameter $f$ may be compared
with the mass scale of the  {\bf $\sigma $} model of mesonic
theory [16]. The confining phase of Dick model is parameterized by
non zero mass for the dilaton and non vanishing  $f$. In trying to
analyze the potential $V_D(r)$ we have observed that the Dick
problem has a perfect formal analogy  with the problem of building
the Eguchi-Hanson metric in 4d N=2 supersymmetric harmonic
superspace. This formal similarity  appears at several levels. In
the introduction we have quoted some of these striking analogies.
For example, vanishing masses for both Dick and Eguchi-Hanson
scalar fields lead to trivial potentials. An other example is that
the mass scale $f$ which is introduced by hand and interpreted as
a compactification scale by Dick plays a similar role as the 4d
N=2 FI coupling appearing in the Eguchi Hanson model (32). Recall
by the way that now it is well established that the FI couplings
are of order of the compactification scale since they are just the
Kahler moduli of the Calabi Yau threefold on which the 10d type II
superstring are compactified  on. To understand the thing behind
the striking similarity between the Dick problem  and the Eguchi
Hanson one, we have reformulated the Dick problem as a one
dimensional field theory. As a consequence we have found a general
formula for the inter-quark potential V(r) which of course depend
on the nature of the dilaton-gluon $G[\phi]$ as shown on eq.(4).
The beauty of this formula is not only because it extends the
Coulomb and Dick theory but also because it can be compared with
known parameterizations of the confinement especially the
contribution of the quark and gluon vacuum condensates. From this
point of view, the Dick model as we have formulated it, may be
viewed as a phenomenological theory modeling the non-perturbative
contributions responsible of confinement. In this regard a more
explicit analysis will be presented in [17].\\Moreover, having at
hand the 1d field theoretical formulation of the Dick model and
the analogy with 4d N=2 Eguchi Hanson model, we have shown how the
axion field may be incorporated in the game in agreement with the
requirement of F-theory and 10d type IIB superstrings according to
which the dilaton and the axion form a complex field. In our
formulation the dilaton-axion model is represented by a 1d U(1)
gauge theory of lagrangian (52). The  $U(1)\approx SO(2)$ symmetry
rotates the dilaton and the axion fields and allows to interpret
the Dick mass scale as a kind of FI coupling.\\
\\
\\
\\
\small This research work has been supported by the program PARS number 372-98 CNR.\\
\\
\\
\newpage
{\bf References}
\begin{enumerate}
\item [[1]] M. Green, J. Schwartz and E. Witten, « Superstring Theory », Cambrige University Press (1987).
\item [[2]] C. Vafa, « Lectures on String and Dualities » HUTP-79/A009, hep-th/9702201.
\item [[3]] A. Sen, « Unification of string dualities », hep-th/9609176.
\item [[4]] C. Vafa, « Evidence for F-theory», HUTP-96/A004, hep-th/9602022.
\item [[5]] N. Seiberg and E. Witten , hep-th/9602022; Nucl. Phys. B426(1994)19 and   Nucl. Phys. B431(1994) 484.
\item [[6]] R. Dick, "Confinment from a massive scalar in QCD", LMU-TPW-98/06, to appear in Eur. Phys. Journal.
C6 (1999). Proceedings of the  Miniworkshop on "Recent Developements in Quantum Field
Theory", UFR-PHE, Rabat (October 1998) to appear.
\item [[7]] R. Dick, "Implication of dilaton in gauge theory and cosmology", hep-th/9701047, LMU-TPW-97/10.
Phys. lett. B397 (1997)193; Phys. lett. B409 (1997)321.
\item [[8]] T. Huang, Z.Huang, Phys. Rev. D39 (1989)1213.
\item [[9]] J.G. Bian, T. Huang and Q.X.Shen, « The quarkonium  potential within  short and intermediate range in QCD», BIHEP-TH-92-38 and references therein.
\item [[10]] S. Katz, P. Mayr and C. Vafa, « Mirror summetry and exact solution of 4d N=2 gauge theories I », HUTP-97/A025, OSU-M.97-5, IASSNS-HEP 97/65, hep-th/9706110.
\item [[11]] A. Galperin, E. Ivanov, V. Ogievetsky and P.K. Townsend, Class. Quantum
.Grav. 3 (1986) 625.
\item [[12]] A. Galperin, E. Ivanov, S. Kalitzin, V. Ogievetsky and E. Sokatchev, Class. Quantum .
Grav. 1 (1985) 469.
\item [[13]] J. Wess and J. Bagger, « Supersymmetry and supergravity », Princeton Press (1983).
S. Ferrara and C. Savoy, Supergravity  81 (Cambridge U.P; Cambridge 1982).
\item [[14]] E.A. Gibbons and P.J. Ruback, Commun. Math. Phys. 115 (1988) 267.
\item [[15]] G.W. Gibbons, D. Olivier, P.J. Ruback and G. Valent,  Nucl . Phys. B296(1988)679.\\
 E. Sahraoui and E.H. Saidi, Class. Quantum. Grav. 16 (1999) 1.
\item [[16]] K. Wehrberger, Phys. Rep. 5-6 (1993) 273.
\item [[17]] M. Chabab, R. Markazi and E.H. Saidi, (in progress).
\item [[18]]M.A. Shifman, A.I. Vainshtein and V.I. Zakharov, Nucl. Phys. B147 (1979) 385.
\end{enumerate}
\end{document}